\documentclass[a4paper,twocolumn, nofootinbib,superscriptaddress,aps,prl,eqsecnum,notitlepage,showkeys]{revtex4-1}
\usepackage{multirow}
\usepackage{graphicx}
\usepackage{bm}
\usepackage{array}
\usepackage{dcolumn}
\usepackage{hyperref}
\usepackage{gensymb}
\usepackage{amsmath}
\usepackage{dcolumn}    
\usepackage{ifpdf}
\usepackage{amssymb,lineno,amsfonts}
\usepackage{graphicx}   
\usepackage{bm}         
\usepackage{bbm}
\usepackage{mathrsfs}
\usepackage{upgreek}
\usepackage{mathtools}
\usepackage{epstopdf}
\usepackage{setspace}
\usepackage{hyperref}
\usepackage{natbib}
\usepackage{multirow}
\usepackage{esvect}
\usepackage{soul}
\usepackage{amsmath}
\usepackage[usenames,dvipsnames]{xcolor}
\definecolor{med-blue}{RGB}{25,25,112}
\hypersetup{colorlinks, linkcolor={blue},citecolor={Blue}, urlcolor={blue}}
\usepackage{times}
\usepackage [english]{babel}
\usepackage [autostyle, english = american]{csquotes}
\MakeOuterQuote{"}

\begin{document}

\title{Investigations of the heterometallic Ludwigite Ni${_2}$AlBO${_5}$}
\author{Jitender Kumar}
\affiliation{Department of Physics, Indian Institute of Science Education and Research \\ Dr. Homi Bhabha Road, Pune, Maharashtra-411008, India}
\author{Deepak John Mukkattukavil}
\affiliation{Department of Physics, Indian Institute of Science Education and Research \\ Dr. Homi Bhabha Road, Pune, Maharashtra-411008, India}
\author{Arpan Bhattacharyya}
\affiliation{Saha Institute of Nuclear Physics, 1/AF Bidhannagar, Kolkata, India }
\author{Sunil Nair}
\affiliation{Department of Physics, Indian Institute of Science Education and Research \\ Dr. Homi Bhabha Road, Pune, Maharashtra-411008, India}
\affiliation{Centre for Energy Science, Indian Institute of Science Education and Research,
	\\ Dr Homi Bhabha Road, Pune, Maharashtra-411008, India}
\date{\today}

\begin{abstract} 
We present magnetic, thermodynamic, dielectric and structural investigations on the aluminoborate Ni${_2}$AlBO${_5}$, belonging to the ludwigite family. Room temperature structural refinement suggests that the system crystallizes in the orthorhombic $Pbam$ symmetry, in similarity with most members of this material class. Magnetic and thermodynamic measurements shows that the system undergoes a phase transition to an antiferromagnetic state at 38K, signatures of which are also seen in the lattice parameters and the dielectric constant. Short range magnetic correlations appear to persist to much higher temperatures - a regime in which the ac susceptibility exhibits a power law temperature dependence, in agreement with that expected for random exchange Heisenberg antiferromagnetic spin chains. 
\end{abstract}
\maketitle
\section{Introduction}
Strongly correlated transition metal oxides in restricted geometries continue to be at the forefront of research in condensed matter physics. The presence of structural sub-units or highly anisotropic magnetic interactions can result in an effective reduction of the dimensionality of these inherently three dimensional entities, which in turn manifests itself in the form of novel physical properties \cite{narlikar}. The anhydrous oxyborates of the form $M{_2}M'$BO${_5}$ (where $M$ and $M'$ are divalent and trivalent metal ions respectively) crystallizing in the \emph{Ludwigite} structure have attracted attention in the recent past for these very reasons \cite{charge, Fe, warw, natural, Sn}. A defining feature of the ludwigite structure is the existence of a zig-zag wall comprising of edge sharing $MO{_6}$ and $M'O{_6}$ octahedra - with these triads running along one of the crystallographic axis (Figure1). The remarkable flexibility of this structure in accommodating a variety of different metal ions in the $M$ and $M'$ positions has also meant that the Ludwigites remain a fertile playground for the exploration of low dimensional magnetism \cite{attfield}. 

\begin{figure}[b]
\centering
\hspace{-0.5cm}
\includegraphics[scale=0.45]{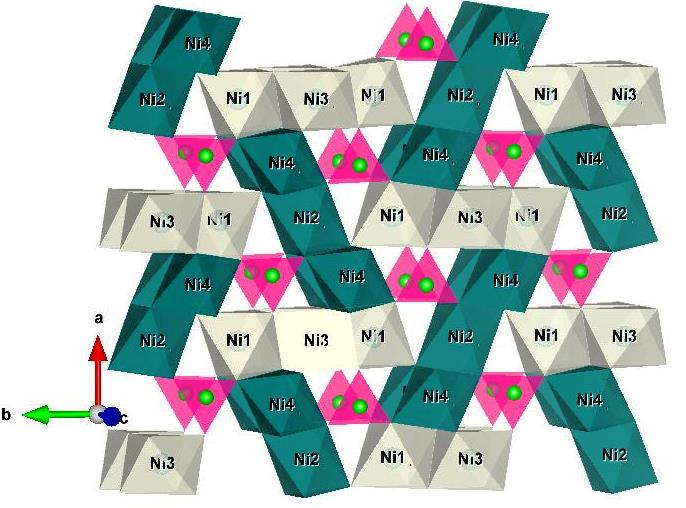}
	\caption{The crystal structure of the Ni$_2$AlBO$_5$, with Ni3-N1-N3 and Ni4-Ni2-Ni4 forming the 3 leg ladder characteristic of the Ludwigite structure. Here, the Boron atoms are depicted in green.}
	\label{Fig1} 
\end{figure}

Broadly classified as homometallic (where $M$ = $M'$) or heterometallic (where $M \neq M'$), the magnetic and electronic properties of the Ludwigites are primarily dictated by which ions occupy the four distinct crystallographic sites available within the zig-zag chain \cite{zigzag}. For instance, the homometallic Fe$_3$BO$_5$ system is reported to exhibit an ordering transition at 283 K, associated with the doubling of the lattice along the crystallographic $c$ axis \cite{charge}. In addition, magnetic transitions are also observed at 110K and 70K \cite{transitions}. On the contrary, the homometallic Co$_3$BO$_5$ does not exhibit this structural transition, and also undergoes a solitary magnetic transition at 45K \cite{Fe, cobolt}, reflecting the diversity of magnetic and electronic states within this family \cite{Fe, div, jittu,Moshkina,Moshkina2}. The heterometallic ludwigites exhibits a staggering array of properties depending on the combination of constituent magnetic (or non-magnetic) metal ions, and phenomena varying from spin and cluster glasses, magneto-dielectricity, charge ordering and multiferroicity have been reported \cite{Ti, Jitender, charge, Martin}.  

Interestingly, though Fe and Co based homo and heterometallic ludwigites have been investigated in some detail in the past,  Ni based systems have hardly been explored. For instance, single phase specimens of homo-metallic Ni${_3}$BO${_5}$ remains  to be reported till date. The only notable exceptions are the \emph{bonaccordite}  Ni${_2}$FeBO${_5}$ and the closely related Ni${_2}$MnBO${_5}$ , which were reported to exhibit a series of magnetic transitions as a function of reducing temperature\cite{Freitas,Moshkina2}. In the former, it was suggested that these transitions were associated with the ordering within the Fe sublattice (at 106K), followed by an ordering in the Ni sublattice (at 46K), with the coupling between both of them manifesting itself in the form of a third transition at 15K \cite{Freitas}.  In the latter, two magnetic transitions (at 85K and 71K) were reported, and was taken to be evidence of loosely coupled magnetic sub-systems ordering separately \cite{Moshkina2}. Here, we report on the hitherto  un-investigated heterometallic ludwigite Ni$_2$AlBO$_5$. We observe a robust para-antiferromagnetic phase transition at around 38K, as inferred from magnetization and specific heat measurements.  Temperature dependent X-ray diffraction measurements rule out the possibility of a structural transition down to 15K, and the system remains in the orthorhombic $Pbam$ symmetry on either side of the magnetic transition. Short range magnetic correlations appear to persist well above the magnetic ordering temperature, and in this regime the ac susceptibility exhibits a power law temperature dependence - a feature associated with one dimensional magnetic chains with random Heisenberg interactions\cite{Dasgupta}. The dielectric constant is seen to exhibit a sharp feature at the magnetic transition, indicating a finite magneto-dielectric coupling in this system. 

\begin{figure}
  \centering
  \hspace{-1cm}
  \includegraphics[scale=0.35]{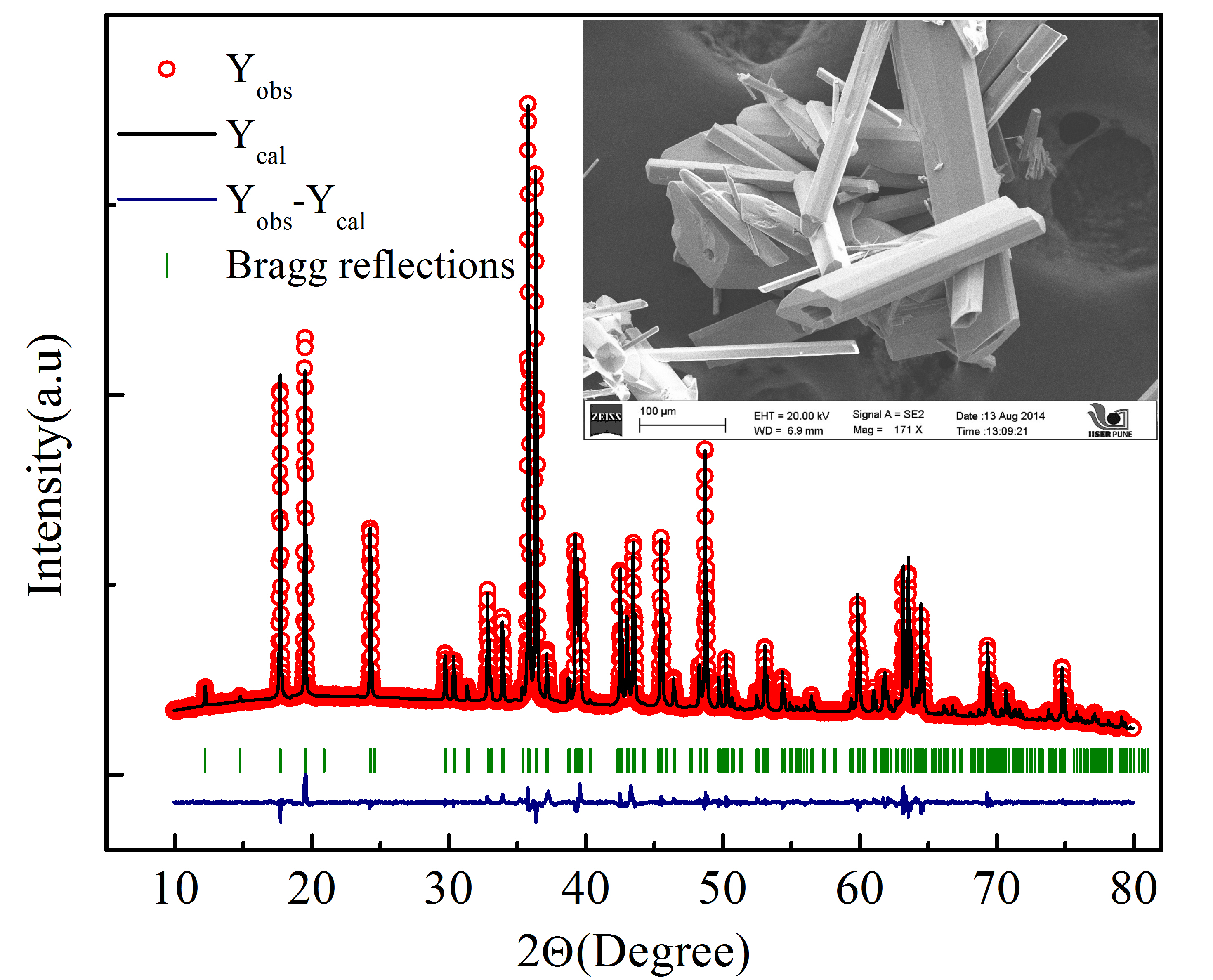}
  \caption{Rietveld refined x-ray diffraction data of the Ni$_2$AlBO$_5$ at room temperature. This corresponds to the fitting parameters of R$_{wp}$=6.45, R$_{exp}$ =3.81 and $\chi$$^2$=2.87. The inset shows a scanning electron micrograph exhibiting both needle-like rods and hollow tube-like crystallites. }
  \label{Fig2}
\end{figure}

\section{Experimental Methods}
Small needle like crystallites of Ni${_2}$AlBO${_5}$ were synthesized using a reactive flux technique. Stoichiometric amounts of NiO and Al$_2$O$_3$ were mixed with excess borax (Na$_2$B$_4$O$_7$ · 10H$_2$O) in the ratio of 1:5, and then ground using a ball mill at 120 rpm for 12 hours to make a fine homogeneous mixture. This mixture was treated at 1000$\degree$C for 90 hours in an alumina crucible followed by slow cooling to 740$\degree$C (at the rate of 5$\degree$C/hour), after which the furnace was turned off. Fine needle-like crystallites of the target material were extracted from the crucible and finally washed using warm dilute HCl to remove the excess borax. Phase purity was confirmed using a Bruker D8 Advance diffractometer with Cu K$\alpha$ radiation. Low-temperature x-ray diffraction was obtained using the powder diffractometer at beamline BL-18B, Photon Factory, KEK, Japan, using an X-ray wavelength of 0.8019$\AA$. Room temperature structural details were analyzed by the Rietveld method using the FULLPROF refinement program, and the variation in the lattice parameters of the low-temperature scans was determined using a Le Bail fit. Specific heat and magnetization measurements were performed using a Quantum Design PPMS and an MPMS-XL SQUID magnetometer respectively. Temperature-dependent dielectric measurements were performed in the standard parallel plate geometry, using a Novo-Control high-performance impedance analyzer using an excitation ac signal of 1 V.

\section {Results and Discussions} 
\begin{table}[hb]
\begin{tabular}{|c|c|c|c|c|c|}
 		\hline
 		\multicolumn{6}{|c|}{Ni$_2$AlBO$_5$} \\
 		\multicolumn{6}{|c|}{Temperature = 296 K} \\
 		\multicolumn{6}{|c|}{Space Group : $Pbam$} \\
 		\multicolumn{6}{|c|}{Crystal system: Orthorhombic }\\
 		\multicolumn{6}{|c|}{ a=9.1055(2) ${\AA}$}\\
 		\multicolumn{6}{|c|}{ b=12.0131(2) ${\AA}$ }\\
 		\multicolumn{6}{|c|}{c= 2.9418(0) ${\AA}$ }\\
 		\multicolumn{6}{|c|}{${\alpha}$ = ${\beta}$ = ${\gamma}$ = 90 \degree} \\
 		\hline
 		\hline
 		Atom & Wyckoff & x/a &y/b & z/c & Occ\\
 		\hline
 		Ni$^1$      & 2b &0& 0 & -0.5 & 0.78\\
 		Al$^1$      & 2b &0& 0 & -0.5 & 0.22\\
 		Ni$^2$      & 2c &0.5& 0 & 0 & 0.57\\
 		Al$^2$      & 2c &0.5& 0 & 0 & 0.43\\
 		Ni$^3$      & 4h &0.99699& 0.28128 & -0.5 & 0.81\\
 		Al$^3$      & 4h &0.99699& 0.28128 & -0.5 & 0.19\\
 		Ni$^4$      & 4g &0.23574& 0.1139 & 0 & 0.30\\
 		Al$^4$      & 4g &0.23574& 0.1139 & 0 & 0.70\\
 		B$^1$      & 4g &0.27409& 0.35773 & 0 & 1\\
 		O$^1$      & 4g &0.34979& 0.46308 & 0 & 1\\
 		O$^2$      & 4h &0.11376& 0.14442 & -0.5 & 1\\
 		O$^3$      & 4g &0.11921& 0.36098 & 0 & 1\\
 		O$^4$      & 4h &0.37747& 0.07701 & -.5 & 1\\
 		O$^5$      & 4g &0.34134&  0.25770 & 0 & 1\\
 		\hline
 	\end{tabular}
 	\caption{Structural Parameters of Ni$_2$AlBO$_5$ as determined from the Rietveld refinement of room temperature X-ray diffraction data.}
 	\label{Table1}
 \end{table}
\begin{figure*}
	\centering
	\hspace{-0.45cm}
	\includegraphics[scale=0.5]{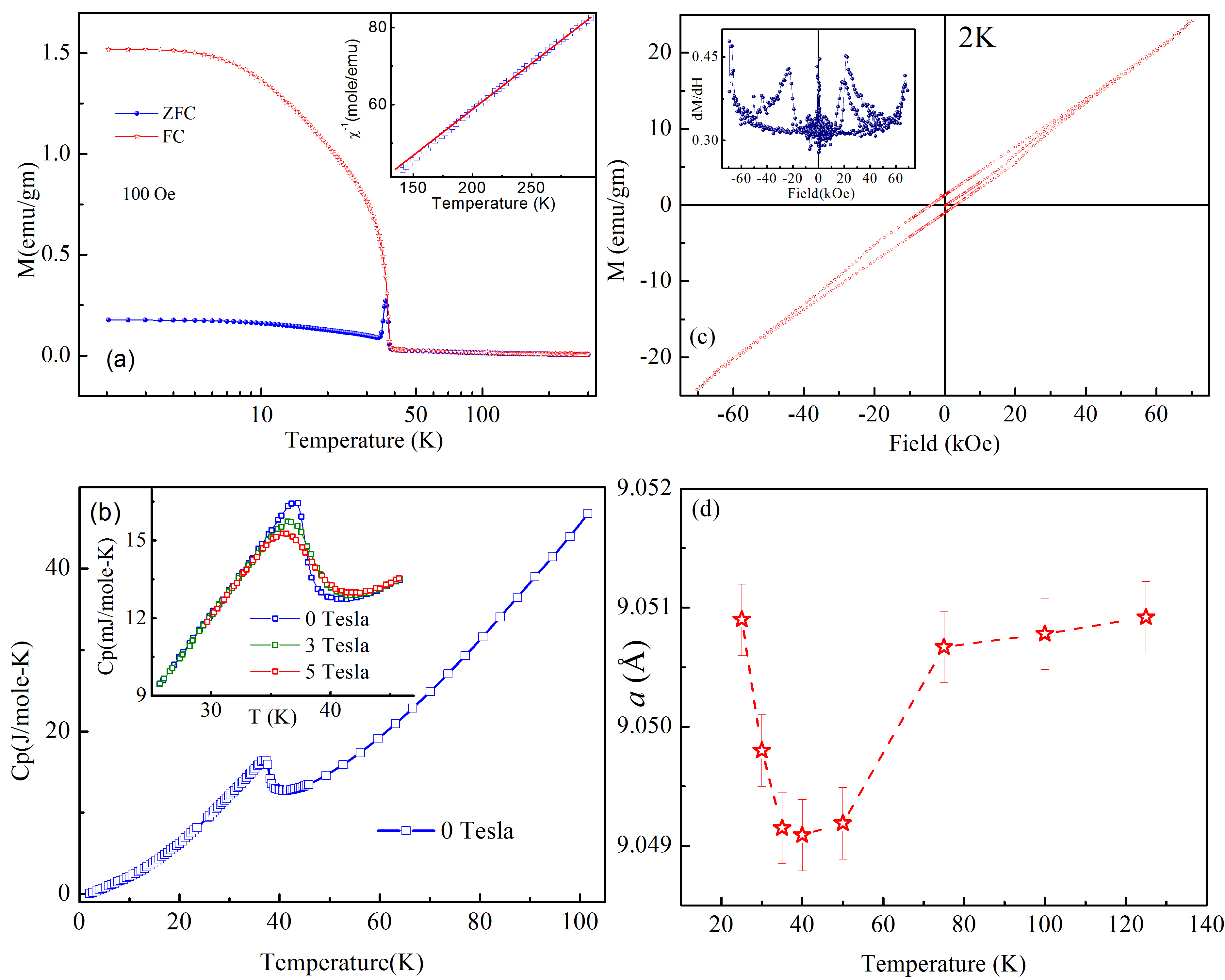}
\caption{(a) DC Magnetization in Ni${_2}$AlBO${_5}$ as measured under 100 Oe applied DC field in the ZFC and FC protocols. The inset depicts the inverse susceptibility fitted with a Curie-Weiss form in the paramagnetic regime (b) Heat capacity measured in zero magnetic field (main panel) exhibiting a sharp anomaly at the magnetic transition temperature. The upper inset shows the heat capacity data taken under 3 and 5 Tesla magnetic fields (c) The magnetization isotherm of Ni$_2$AlBO$_5$ as measured at 2 K, exhibiting a metamagnetic transition. In the inset, a derivative of M versus H is shown (d) The variation of the lattice parameter ${a}$ as a function of temperature, exhibiting a clear change in the vicinity of the magnetic transition.}
\label{Fig3}
\end{figure*}

The Rietveld refinement of the powder x-ray diffraction data as obtained on powdered Ni$_2$AlBO$_5$ specimens is shown in Figure (2). The absence of any impurity peaks indicates the single phase nature of the as-prepared sample. Scanning electron micrographs reveals that the samples are largely composed of solid needle-like crystallites and a few hollow tube like structures as is shown in the inset. We note that there are no reports of such hollow tubular morphologies in this family of systems, and this aspect probably deserves more careful attention. Elemental analysis using Energy Dispersive x-ray analysis suggest that Ni:Al ratio is 1.95:1.05. The analysis of x-ray diffraction data reveals that Ni$_2$AlBO$_5$ crystallizes in the orthorhombic $Pbam$ structure, and the structural parameters derived from the Rietveld refinement is summarized in Table 1.  Our data indicates that all the four crystallographic sites available for the magnetic Ni species are diluted by Al, with Ni$^4$ and Ni$^3$ being the most preferred and least preferred sites respectively. The chemical composition as deduced from the Rietveld analysis is Ni${_{1.8}}$Al${_{1.2}}$BO${_5}$, which is in reasonable agreement with both the targeted composition and EDAX data. 
 
Figure 3 summarizes the physical property measurements on the Ni${_2}$AlBO${_5}$ system.The temperature dependent magnetization as measured in the standard zero field cooled (ZFC) and field cooled (FC) protocols is depicted in Figure 3a, where a sharp feature at around 38K is clearly seen. In contrast, the Ni sublattice in the closely related Ni$_2$FeBO$_5$ system is reported to order at around 42 K\cite{REHAC}, suggesting that the dilution of the magnetic lattice in the case of Ni${_2}$AlBO${_5}$ does not appear to dramatically decrease the temperature at which long range magnetic order sets in. This onset of magnetic order is also characterized by a sharp anomaly in the heat capacity as is shown in Figure.3b.  This feature is robust against the application of magnetic fields, which shifts slightly to lower temperatures as a function of increasing magnetic field (inset of Figure 3b) as is expected for antiferromagnetic systems. A Curie-Weiss fit to the high temperature magnetization data (inset of Figure 3a) was seen to be valid only above 200K, indicating that short range magnetic correlations are present much above the magnetic ordering temperature. This fit yields an effective magnetic moment of $2.87\mu{_B}$ which is very close to the theoretical value of 2.828 $\mu{_B}$ expected for Ni$^{2+}$.  The intercept gives a $\theta_{CW}$ value of -45.86K, indicating the antiferromagnetic nature of magnetic correlations.  In contrast, the closely related aluminoborate Co${_2}$AlBO${_5}$ had yielded a $\theta_{CW}$ value of -9.1 K, though the antiferromagnetic transition temperature was similar ($T{_N}\approx$ 42K) - presumably as a consequence of mixed ferro and antiferromagnetic interactions\cite{Jitender}. This was also reflected in the observation of a reentrant superspin glass state in the Co aluminoborate, whereas there is no trace of such a glassy state in the Ni system. Deep within the superspin glass state of Co${_2}$AlBO${_5}$,  sharp magnetization steps, presumably arising due to the catastrophic evolution of the antiferromagnetic-ferromagnetic phase boundary was also observed. In contrast, the $MH$ isotherms as measured in Ni${_2}$AlBO${_5}$ is in line with that expected for a normal antiferromagnetic specimen, as is seen in Figure 3c. Though a meta-magnetic transition is seen at about 2 Tesla, this transition appears to be associated with a routine (spin flop like) rotation of the antiferromagnetic sublattice magnetization as a function of the applied field \cite{Jitender}. Incidentally, the homo-metallic Fe$_3$BO$_5$ ludwigite was reported to exhibit a structural transition around room temperature due to the charge ordering of Fe$^{2+}$ ions \cite{charge}. To examine the possibility of a similar structural transition in Ni$_2$AlBO$_5$, we have performed low-temperature x-ray diffraction measurements down to 20K. Our observations confirms the absence of such a structural transition and the system appears to be remain in the orthorhombic $Pbam$ symmetry down to the lowest measured temperatures. However, in similarity to the Co aluminoborate, we observe a clear anomaly in the lattice parameter ${a}$ at the antiferromagnetic ordering temperature as is depicted in Figure 3d. 
\begin{figure}
	\centering
	\hspace{-0.5cm}
	\includegraphics[scale=0.3]{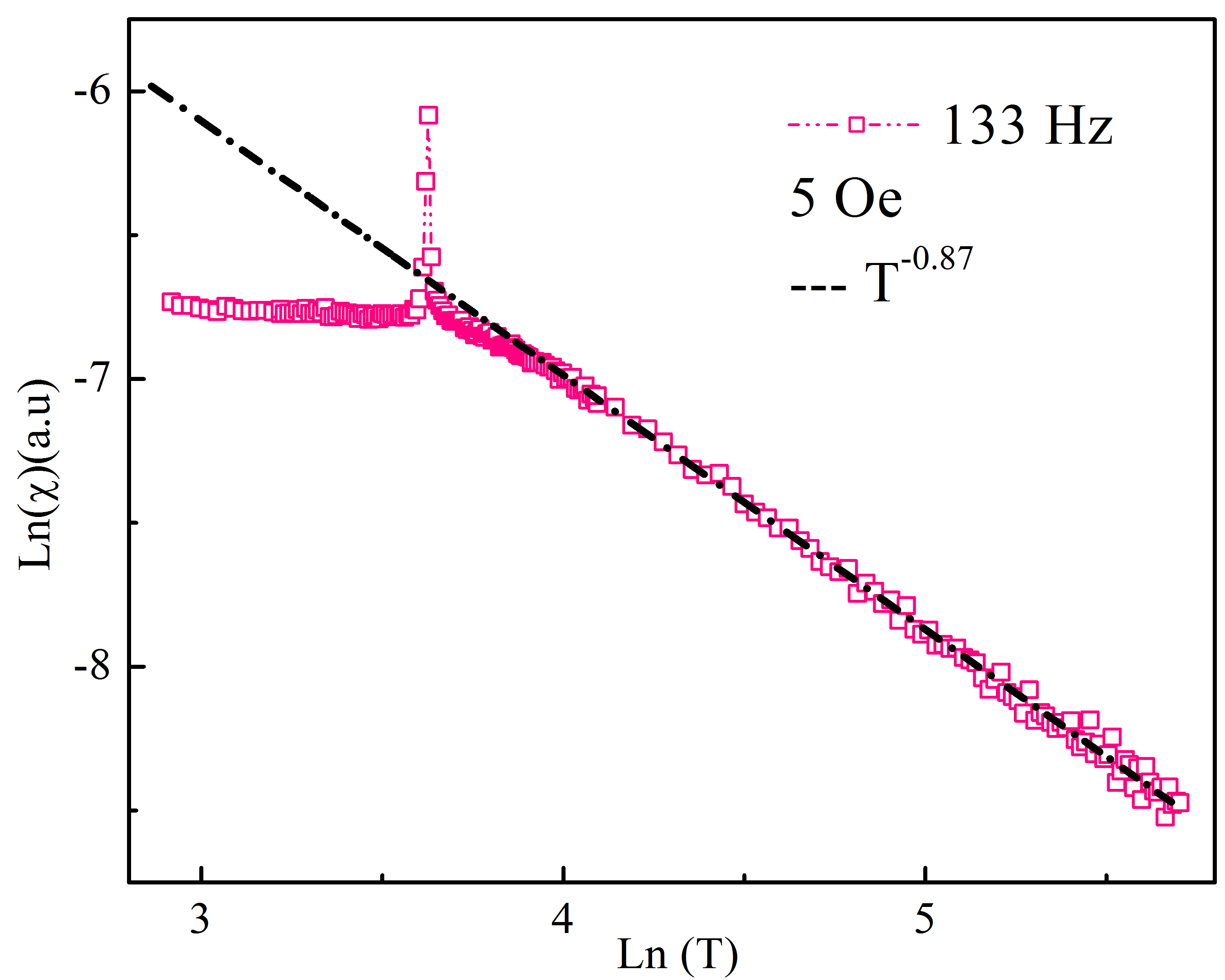} 
	\caption{Log-log plot of the real part of ac susceptibility versus temperature as measured in Ni${_2}$AlBO${_5}$, with the straight depicting a power law fit to the data.   }
	\label{Fig4 }
\end{figure}
We note that magnetic correlations in Ni${_2}$AlBO${_5}$ appear to extend to temperatures well above $T{_N}$ - presumably a consequence of the low dimensional nature of the magnetic interactions in this system. Interestingly, the temperature dependence of ${\chi}_{ac}$ in this region above $T{_N}$ is well described by a power law ${\chi}(T)$=$T^-$$^{\alpha}$, (with ${\alpha}$ = 0.87) as is  evident in the log-log plot depicted in Figure(4). Such a power law behavior of the magnetic susceptibility is a known characteristic of Random Exchange Heisenberg Antiferromagnetic Chains (REHAC)\cite{Dasgupta, Vanadium}, and has been observed earlier in the ludwigite Ni$_2$FeBO$_5$\cite{REHAC}, and some closely related Warwickite systems \cite{MgTO, Guimar,Vanadium}.  Initially formulated to explain the magnetic behavior of (S = 1/2)  Heisenberg spins on a chain, this anomalous temperature dependence of the magnetic susceptibility has now been in observed in both organic and inorganic systems, where the crystallographic structure effectively reduces the dimensionality of the magnetic interactions to be quasi 1D like\cite{PhysicaB,TCNQ,Vanadium,qui}. A defining feature of the ludwigite family is the presence of zig-zag chains of edge sharing octahedra, with Ni${^{2+}}$ and Al${^{3+}}$ occupying the central metal ion position. Though the three legged triads in the ludwigites are essentially 2 dimensional objects, the nearly random substitution of the non-magnetic Al$^{3+}$ in the four crystallographic positions coupled with weaker interactions between these triads mimics the scenario in the Random Exchange Heisenberg Antiferromagnetic Chains\cite{Vanadium, PhysRevB}. This gives rise to an extended region in temperature, where the susceptibility varies as $T^-$$^{\alpha}$. It appears that on further reduction in temperature, the inter-triad coupling becomes increasingly dominant, and finally culminates in a global (3D) antiferromagnetic order at 38K. 

\begin{figure}
 	\centering
 	\hspace{-0.5cm}
 	\includegraphics[scale=0.3]{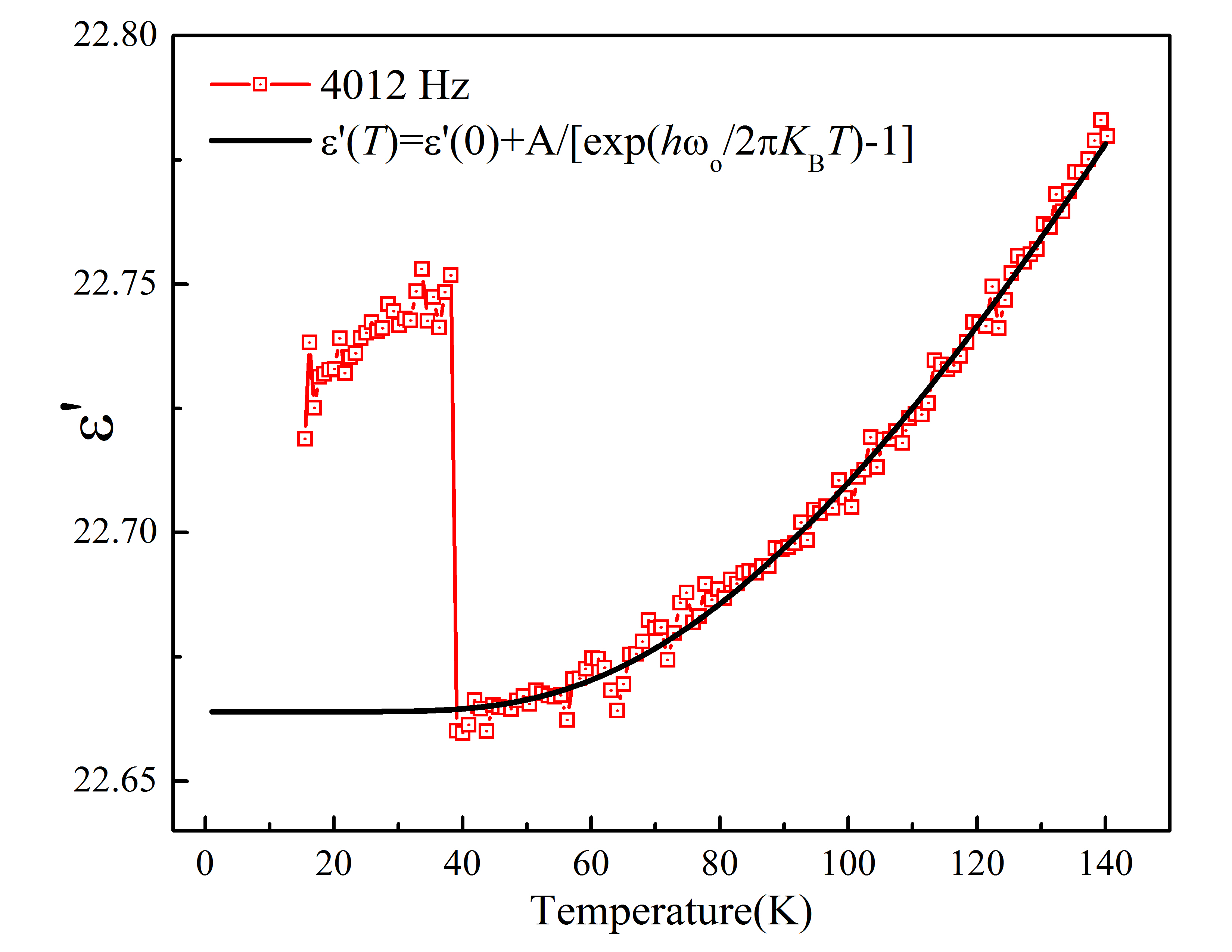}
 	\caption{Real part of dielectric permittivity as a function of temperature. The solid line depicts a fit using the modified Barrett formula (see main text) incorporating the contribution from the lattice dielectric constant alone.}
 	\label{Fig5}
 \end{figure}	

The dielectric properties of some oxyborates have been investigated in the past, and a strong coupling between the magnetic and polar properties have been observed earlier in Fe${_3}$BO${_5}$, Fe${_2}$MnBO${_5}$ and Co${_2}$AlBO${_5}$\cite{Martin, Jitender}. The real part of the dielectric susceptibility as measured on a pressed pellet of Ni$_2$AlBO$_5$ exhibits a sharp anomaly coinciding with the onset of magnetic order as depicted in Figure 5, indicating that the magnetic and dielectric properties are strongly coupled in this system.  Typically, the magnetic and dielectric properties are mediated by spin-phonon coupling through 
$\epsilon^{\prime}(T,H)$= $\epsilon^{\prime}(T)$[1+$\eta$$_{i,j}(S_{i}S_{j})]$
where $\epsilon^{\prime}(T)$ is the temperature dependent lattice dielectric constant in the absence any magnetic ordering or negligible spin-spin correlations and $\eta$ is the spin and dielectric coupling coefficient \cite{LAWES}. The lattice dielectric constant of insulators predominantly depend on phonons lying in the optical range, which in turn are related by the Lyddane-Sachs-Teller relation
$\epsilon^{\prime}_{0}$ =[$\omega^2_L$/$\omega^2_T$]*$\epsilon^{\prime}_{\infty}$
where $\omega_L$, $\omega_T$, $\epsilon^{\prime}_{\infty}$ and $\epsilon^{\prime}_{0}$ are the longitudinal, transverse optical-phonon mode frequencies, the optical dielectric constant and the static dielectric constant respectively. In the previous reports on BaMnF$_4$\cite{Scott},MnF$_2$\cite{MnF2} and MnO\cite{MnO} the temperature dependence of the lattice dielectric constant has been fitted by modified Barrett equation of the form
 $\epsilon^{\prime}_{T}$ =$\epsilon^{\prime}(0)$+$A$/[exp($\hbar$$\omega_0$/k$_B$T)-1]
	where $\epsilon^{\prime}(0)$ is the expected dielectric constant at 0K,  $A$ is a coupling constant and $\omega_0$ is the lowest mean frequency of the final states in the optical phonon branch. A fit using this modified Barrett formalism is indicated by a solid line in Figure(5), yielding $\epsilon^{\prime}(0)$=22.664($\pm0.0008$), $A$=0.80($\pm0.06$), $\omega_0$=202($\pm6$)cm$^{-1}$. It would be interesting to perform Raman or infrared measurements on this system to evaluate if any of the observed modes corresponds to the value which we have deduced from our fitting procedure.  A coupling between dielectric and magnetic properties is also evident from the fact that the dielectric constant exhibits a sharp anomaly at the magnetic ordering temperature.  We note that in contrast to what we observe, it is more common to observe a decrease in the dielectric constant at the onset of magnetic ordering, as has been reported in systems like YMnO$_3$, MnO, SeCuO$_3$ and BiMnO$_3$ \cite{Takagi, MnO, PhysRevLett.,Tokura}. A notable exception is the system Cu$_2$OSeO$_3$, where an enhancement in the dielectric constant at the magnetic transition was reported and attributed due to the complex evolution of magnetic order parameter \cite{enhance}. Additional investigations - especially of the temperature dependent evolution of the magnetic structure - would be required to understand the origin of this feature in Ni$_2$AlBO$_5$. 

\section{Conclusion}
In conclusion, we have investigated the structural, magnetic, thermodynamic and dielectric properties of the hetero-ludwigite Ni$_2$AlBO$_5$. This system is seen to crystallize in an orthorhombic $Pbam$ symmetry and exhibits a long range antiferromagnetic transition at 38 K.   A clear lambda-like phase transition in heat capacity reinforces the true thermodynamic nature of the para-antiferromagnetic phase transition. Low temperature $MH$ isotherms reveal the presence of a metamagnetic transition at 2.2 Tesla. Short range magnetic correlations appears to extend to temperatures far in excess of the magnetic transition temperature, and in this region, the ac susceptibility follows a power law dependence characteristic of random exchange Heisenberg antiferromagnetic chains  A clear discontinuity is observed in the dielectric constant at the magnetic phase transition,  indicating the presence of a  strong magneto-dielectric coupling in the system.   
  
\section{Acknowledgements}
The authors thank S. Panja and S. Singh for help in magnetic and heat capacity measurements. J.K. acknowledges DST India for support through PDF/2016/000911.NPDF. S.N. acknowledges DST India for support through grant no. SB/S2/CMP-048/2013. Authors thank the Department of Science and Technology, India for financial support during the experiments at the Indian Beamline, PF, KEK, Japan. 

\bibliography{Ludwigite_Draft}
 
\end{document}